\documentclass[%
 preprint,
nofootinbib,
 amsmath,amssymb,
 aps,
prl,
 longbibliography,
 lengthcheck,%
]{revtex4-1}

\usepackage{graphicx}% Include figure files
\usepackage{dcolumn}% Align table columns on decimal point
\usepackage{bm}% bold math
\usepackage{hyperref}% add hypertext capabilities
\usepackage{xcolor}
\usepackage{scrextend}
\usepackage{dsfont}

\definecolor{mynicegreen}{RGB}{28,161,85}

\def\ba{\begin{equation}\begin{array}{c}}
\def\ea{\end{array}\end{equation}}
\def\be{\ba\displaystyle}
\def\ee{\ea}

\newcommand{\tr}{{\rm tr}}

\renewcommand{\Pi}{\hat P}

\begin{document}

\preprint{APS/123-QED}

\title{
Quantum speed limit for thermal states
}
% Force line breaks with \\
%\thanks{A footnote to the article title}%

\author{Nikolai Il`in$^{1}$}
\author{Oleg Lychkovskiy$^{1,2,3}$}

\affiliation{$^1$ Skolkovo Institute of Science and Technology,
 Bolshoy Boulevard 30, bld. 1, Moscow 121205, Russia}
\affiliation{$^2$ Department of Mathematical Methods for Quantum Technologies, Steklov Mathematical Institute of Russian Academy of Sciences,
8 Gubkina St., Moscow 119991, Russia}
\affiliation{$^3$ Laboratory for the Physics of Complex Quantum Systems, Moscow Institute of Physics and Technology, Institutsky per. 9, Dolgoprudny, Moscow  region,  141700, Russia}

%\collaboration{CLEO Collaboration}%\noaffiliation

\date{\today}% It is always \today, today,
             %  but any date may be explicitly specified

\begin{abstract}
Quantum speed limits are rigorous estimates on how fast a  state of a quantum system can depart from the initial state in the course of quantum evolution. Most known quantum speed limits, including the celebrated Mandelstam-Tamm and Margolus-Levitin ones,  are general bounds applicable to arbitrary initial states. However, when applied to mixed states of many-body systems, they, as a rule, dramatically overestimate the speed of quantum evolution and fail to provide meaningful bounds in the thermodynamic limit.  Here we derive a quantum speed limit for a closed system initially prepared in a thermal state and evolving under a time-dependent Hamiltonian. This quantum speed limit  exploits  the structure of the thermal state and, in particular,  explicitly depends on the temperature. In a broad class of many-body setups it proves to be  drastically stronger than general quantum speed limits.
\end{abstract}

%\pacs{Valid PACS appear here}% PACS, the Physics and Astronomy
                             % Classification Scheme.
%\keywords{Suggested keywords}%Use showkeys class option if keyword
                              %display desired
\maketitle

%\tableofcontents

\noindent  {\it Introduction.}~~Quantum speed limits (QSL) are a family of fundamental results in quantum mechanics limiting the maximal possible speed of quantum evolution. The first QSL was derived by Mandelstam and Tamm in 1945 \cite{mandelstam1945uncertainty} in a successful attempt to put  the time-energy uncertainty relation on a rigorous basis (see also \cite{fleming1973unitarity}). Decades later, a quite different QSL was derived by Margolus and Levitin \cite{margolus1998maximum}. Further developments included generalizations to mixed states  \cite{uhlmann1992energy,giovannetti2003quantum,frowis2012kind,deffner2013quantum,Mondal_2016_quantum,Campaioli_2018_Tightening}, time-dependent Hamiltonians \cite{uhlmann1992energy,braunstein1994statistical,pfeifer1993fast,Huang_2020_Instantaneous}, open quantum systems \cite{delCampo2013quantum,taddei2013quantum,deffner2013quantum} etc, as reviewed e.g. in \cite{pfeifer1995generalized,frey2016quantum,deffner2017quantum}.   The scope of QSL was broadened to optimal control theory \cite{caneva2009optimal}, quantum resource theory \cite{campaioli2020resource}, abstract quantum information theory \cite{deffner2020quantum}, semiclassical and classical dynamics \cite{Shanahan_2018,Okuyama_2018} etc. Apart from the foundational importance {\it per se}, quantum speed limits enjoy a diverse range of applications, from a deep interrelation between QSLs, orthogonality catastrophe, adiabatic conditions and adiabatic quantum computation \cite{lychkovskiy2017time,il'in2020quantum,lychkovskiy2018necessaryJRLR,kieu2019class,suzuki2019performance,fogarty2020orthogonality}   to ultimate limits for performance of quantum gates \cite{margolus1998maximum,lloyd2000ultimate,svozil2005maximum,santos2015superadiabatic}, quantum heat machines \cite{del2014more,abah2017energy}, quantum transport~\cite{lam2020demonstration} and even quantum batteries \cite{Campaioli2017enhancing,allan2020time}.

Quantum speed limits can be particular useful in many-body systems -- there the exact calculations of the time-dependent state $\rho_t$ is, in general, of prohibitive complexity, and one may hope that QSLs would deliver valuable information on the dynamics not accessible otherwise. It turns out, however, that both the Mandelstam-Tamm (MT) and Margolus-Levitin (ML) QSLs applied to many-body systems often prove to be  notoriously loose \cite{Bukov_2019_geometric}.

Here we derive a QSL with a drastically improved  many-body performance. It applies to a ubiquitous situation when a system is initially prepared in a thermal state of some Hamiltonian $H_0$ and then evolves under a different (possibly, time-dependent) Hamiltonian $H_0+V_t$. We demonstrate the superiority of the derived QSL by comparing it to other paradigmatic QSLs. The results of this comparison are summarised in Table \ref{table}.

The rest of the paper is organized as follows. We first introduce a quantum quench with the system initialised in a thermal state with $V_t=V$ independent on time. Then we discuss figures of merits appropriate to distinguish many-body mixed states. After that  we formulate our QSL for a quench, eq. \eqref{T-QSL}. Then we contrast the performance of our QSL to that of the MT and ML QSLs, as well as to a notable recent QSL by Mondal, Datta and Sazim (MDS)~\cite{Mondal_2016_quantum}. We argue that our QSL has a dramatic advantage over these QSLs in broad classes of many-body setups,  and demonstrate these advantages in two exemplary systems - a spin-boson model and a mobile impurity model.  Then we generalize our QSL to time-dependent potentials, see eq. \eqref{T-QSL general}, and  provide the proof thereof. Finally, we conclude the paper by summarizing the results.

\medskip
\noindent  {\it Quenching a thermal state.}~~ Let us consider a closed quantum system with the Hamiltonian quenched from $H_0$ at $t\leq 0$ to $H_0+V$ at $t>0$. Before the quench the system is in the thermal  state
\be\label{thermal state}
\rho_0=e^{-\beta H_0}/Z_0,\quad Z_0=\tr e^{-\beta H_0},
\ee
$\beta$ being the inverse temperature. After the quench the state of the system $\rho_t$ starts to evolve according to the von Neumann equation
\be\label{Schrodiger equation}
i\partial_t \rho_t = [H_0+V,\rho_t].
\ee
We will refer to $V$ as {\it perturbation}.\footnote{Note that we do not imply that  $V$ is small or treated perturbatively.}

%When we focus on the many-body systems, the properties of $V$ in the thermodynamic limit matter. We will refer to $V$  as {\it local} perturbation whenever $\langle V^2 \rangle_\beta$ is finite in the thermodynamic limit (here and in what follows thermal averaging is understood with respect to $\rho_0$, i.e. $\langle A \rangle_\beta\equiv\tr (\rho_0 A)$ for an arbitrary operator $A$), and {\it finitely disturbing} whenever  $\langle [H_0,V]^2 \rangle_\beta$ is finite in the thermodynamic limit.

\medskip
\noindent  {\it Quantum state distinguishably measures.}~~ Our goal is to assess how far $\rho_t$ can  depart from $\rho_0$.  The difference between two arbitrary mixed quantum states, $\rho_1$ and $\rho_2$, can be quantified by various distinguishably measures, two popular ones being the trace distance
\be
D_{\rm tr}(\rho_1,\rho_2)\equiv (1/2)\,\tr |\rho_2-\rho_1 |
\ee
and the Bures angle
\be
\mathcal L (\rho_1,\rho_2)\equiv \arccos \tr \sqrt{\sqrt{\rho_1} \rho_2 \sqrt{\rho_1}}.
\ee
%where
%$
%F\equiv \left(\tr \sqrt{\sqrt{\rho_1} \rho_2 \sqrt{\rho_1}}\right)^2
%$
%is known as quantum fidelity.
We will mostly employ the trace distance, which is known to have a well-defined operational meaning \cite{helstrom1969quantum,holevo1972quasiequivalence,holevo1973statistical,wilde2013quantum} and can be used for quantum state discrimination in the many-body case~\cite{Markham2008}. In addition, we will provide a QSL in terms of the Bures angle. To prove our QSL we will need a yet different distinguishably measure given by
\be\label{D definition}
D(\rho_1,\rho_2)\equiv  1-\tr(\sqrt{\rho_1}\sqrt{\rho_2}).
\ee
In fact, the three distinguishably measures introduced above are all mutually related by rather strong two-sided inequalities \cite{holevo1972quasiequivalence,audenaert_2014_comparisons} and therefore can be used essentially interchangeably. The two particular inequalities we will use read \cite{holevo1972quasiequivalence}
\begin{align}\label{Holevo inequality}
 D_{\rm tr}\left(\rho_1,\rho_2\right)& \leq \sqrt{D(\rho_1,\rho_2)\big(2-D(\rho_1,\rho_2)\big)} \nonumber \\
  &\leq \sqrt{2 D(\rho_1,\rho_2)}
\end{align}
and \cite{audenaert_2014_comparisons}
\begin{align}\label{Audenaert inequality}
\mathcal L \left(\rho_1,\rho_2\right) & \leq \arcsin \sqrt{D(\rho_1,\rho_2)\big(2-D(\rho_1,\rho_2)\big)}\nonumber \\
 & \leq \arcsin  \sqrt{2 D(\rho_1,\rho_2)}.
\end{align}

As a side remark, we note that the (more easily computable) Hilbert-Schmidt distance is unsuitable for  fairly discriminating many-body states ~\cite{Markham2008} and thus will not be used.

\medskip
\noindent  {\it QSL for thermal states.}~~ The central result of the present paper is the quantum speed limit for thermal states (T-QSL). It reads as follows:
\be\label{T-QSL}
{\textrm{T-QSL:}}\qquad D_{\rm tr}(\rho_0,\rho_t)\leq \sqrt{\beta t} \, \sqrt[4]{-2\,\langle [H_0,V]^2 \rangle_\beta}.
\ee
Here and in what follows thermal averaging is understood with respect to $\rho_0$, i.e. $\langle A \rangle_\beta\equiv\tr (\rho_0 A)$ for an arbitrary operator $A$.

Essentially the same result can be cast in terms of the Bures angle:
\be\label{T-QSL Bures angle}
\mathcal L (\rho_0,\rho_t)\leq \arcsin \left( \sqrt{\beta t} \, \sqrt[4]{-2\,\langle [H_0,V]^2\rangle_\beta} \right).
\ee

The QSLs \eqref{T-QSL} and \eqref{T-QSL Bures angle} follow from more general QSLs \eqref{T-QSL general} and \eqref{T-QSL Bures angle general}, respectively, that are reported and proven in what follows. However, before we turn to the proof, we elucidate the significance and merits of the T-QSL \eqref{T-QSL}.

\begin{table*}
\caption{\label{table} Merits of the thermal QSL \eqref{T-QSL} as compared to three other QSLs, eqs. \eqref{MT QSL}--\eqref{MDS QSL}. {\it Loose} performance implies that the corresponding bound on $ D_{\rm tr}(\rho_0,\rho_t)$ diverges as $\sqrt{N}$ or faster in the thermodynamic limit despite  $ D_{\rm tr}(\rho_0,\rho_t)$ itself being finite. {\it Tight} performance implies the absence of such spurious divergence. The first two lines correspond to the trivial dynamics $\rho_t=\rho_0$. The last two lines correspond to nontrivial dynamics where the complexity of calculating $\rho_t$ becomes  prohibitive for large system sizes. The original form \eqref{MDS QSL original} of the MDS inequality is used in the first two lines, and the modified one \eqref{MDS QSL} -- in the last two lines. }
\begin{ruledtabular}
\begin{tabular}{lcccc}
 &\multicolumn{4}{c}{quantum speed limits}\\
& Mandelstam-Tamm & Margolus-Levitin &  Mondal-Datta-Sazim &thermal \\ \hline
 zero temperature %$\beta=\infty$ 
 & {\bf \textcolor{red}{loose}} & {\bf \textcolor{red}{loose}} & {\bf \textcolor{mynicegreen}{exact}} &  {\bf \textcolor{mynicegreen}{exact}}\\[1em]
trivial perturbation% $[H_0,V_t]=0$ 
& {\bf \textcolor{red}{loose}} & {\bf \textcolor{red}{loose}} & {\bf \textcolor{mynicegreen}{exact}} &  {\bf \textcolor{mynicegreen}{exact}}\\[1em]
 local perturbation %$\langle V_t^2 \rangle_\beta=O(1)$ 
& {\bf \textcolor{red}{loose}} & {\bf \textcolor{red}{loose}} &  {\bf \textcolor{mynicegreen}{tight}} &  {\bf \textcolor{mynicegreen}{tight}}\\[1em]
finitely disturbing nonlocal perturbation &  {\bf \textcolor{red}{loose}} &  {\bf \textcolor{red}{loose}} & {\bf \textcolor{red}{loose}} &  {\bf {\bf \textcolor{mynicegreen}{tight}}}
%\\
% perturbation: $\langle [H_0,V_t]^2 \rangle_\beta=O(1)$, $\langle V^2 \rangle_\beta\rightarrow\infty$ &   &  &  &
%expressed via a local &  {\bf \textcolor{mynicegreen}{yes}} &  {\bf \textcolor{mynicegreen}{yes}} & {\bf \textcolor{red}{no}} &  {\bf \textcolor{mynicegreen}{yes}}\\
%physical observable  & & & &
\end{tabular}
\end{ruledtabular}
\end{table*}

\medskip
\noindent  {\it General QSLs.}~~
We would like to discuss the merits of the T-QSL \eqref{T-QSL} in comparison with three general (i.e. applicable to arbitrary, not necessarily thermal initial states) QSLs. The first two are the MT~\cite{mandelstam1945uncertainty,fleming1973unitarity,uhlmann1992energy} and ML~\cite{margolus1998maximum,giovannetti2003quantum,deffner2013quantum}  QSLs which read
\begin{align}
{\textrm{MT QSL:}}~~ & D_{\rm tr}(\rho_0,\rho_t)\leq \Delta E\, t, \label{MT QSL} \\
 & \qquad \Delta E \equiv  \sqrt{\langle (H_0+V)^2 \rangle_\beta-\langle H_0+V \rangle_\beta^2}, \nonumber\\
{\textrm{ML QSL:}}~~ & D_{\rm tr}(\rho_0,\rho_t)\leq \sqrt{2 \, \overline E \,t}, \label{ML QSL}\\
& \qquad  \overline E  \equiv  \, \langle H_0+V \rangle_\beta-E_{\rm gs}.  \nonumber
\end{align}
Here $\Delta E$ is the quantum uncertainty of the Hamiltonian $H_0+V$ in the thermal state \eqref{thermal state}, $E_{\rm gs}$ is the ground state energy of this Hamiltonian and $ \overline E$ is the average energy relative to $E_{\rm gs}$.

Both the Mandelstam-Tamm (MT) and Margolus-Levitin (ML) QSLs are saturated by a particular pure state~\cite{levitin2009fundamental}. However, this state is very special: it is a coherent, equally weighted superpositions  of  two lowest eigenstates of the Hamiltonian.

When it comes to thermal states, both $\Delta E$ and $ \sqrt{\overline E}$  scale as $\sqrt{N}$ in the thermodynamic limit ($N$ being the number of particles which grows with the particle density kept constant). Therefore, as noted in ref. \cite{Bukov_2019_geometric}, the MT and ML QSLs misleadingly suggest that a state can always evolve into an (almost) orthogonal one in no time. We will see that, in fact, this $\sqrt{N}$ divergence is spurious for a broad class of perturbations, and MT and ML QSLs dramatically overestimate the speed of quantum evolution at a finite temperature.

The third QSL we have picked for comparison is the  MDS QSL. In its original form it reads
\be\label{MDS QSL original}
D(\rho_0,\rho_t)\leq 2 \left(\sin\left(t\,\sqrt{-{\rm tr} \,[\sqrt{\rho_0},V]^2}/2\right)\right)^2,
\ee
valid for times such that the argument of $\sin$ does not exceed $\pi/4$)~\cite{Mondal_2016_quantum}. However, the quantity ${\rm tr} \,[\sqrt{\rho_0},V]^2$, known as Wigner-Yanase skew information (with respect to the observable $V$) \cite{Wigner_1963}, is hardly computable in the many-body case. To obtain a practical bound, we employ the inequality $-{\rm tr} \,[\sqrt{\rho_0},V]^2\leq 2 \langle V^2 \rangle_\beta$. This way we reduce the original MDS QSL to a simpler bound, which we present in terms of trace distance (with the help of inequality \eqref{Holevo inequality}):
\begin{equation}\label{MDS QSL}
{\textrm{MDS QSL:}}\qquad D_{\rm tr}(\rho_0,\rho_t)\leq t \sqrt{2 \langle V^2 \rangle_\beta}.
\end{equation}
In what follows we will refer to this modified MDS bound simply as the MDS QSL.

% Note that the MDS QSL becomes similar to the T-QSL in the high-temperature limit. Indeed, if we approximate $\rho_0\simeq (\mathds{1}-\beta H_0)/Z_0$, we get from eq. \eqref{MDS QSL} an approximate bound $D_{\rm tr}(\rho_0,\rho_t)\lesssim \sqrt{\beta t} \, \sqrt[4]{-(1/2)\,\langle [H_0,V]^2 \rangle_\beta}$, that differs from the T-QSL \eqref{T-QSL} only by the factor $1/\sqrt2$.

At this point it is worth noting that all QSLs considered in the present paper (except the bound \eqref{MDS QSL original}) can be regarded as inequalities relating nonequilibrium dynamics (left hand side) to an equilibrium expectation value of some physical observable or a function thereof (right hand side). This type of QSLs is particularly useful in the many-body setting since the equilibrium expectation values are more easily accessible both theoretically and experimentally than the far-from-equilibrium dynamics.  Different types of QSLs  (see e.g. \cite{Campaioli_2018_Tightening}) are not discussed here.

%This physics in equilibrium  depends on an equilibrium value of the concrete physical observable,  $-2[H_0,V]^2$, in contrast to the MDS QSL. This way, T-QSL relates nonequilibrium dynamics to the  physics in equilibrium -- a feature absent in the MDS QSL. We note that a similar problem plagues  a QSL derived in ref. \cite{Campaioli_2018_Tightening}, which depends on the quantity $\tr \big(\rho_0^2 \,V- (\rho_0 V)^2\big)$ that is not a physical observable and can not be efficiently computed in the many-body case.

Now we are in a position to compare the general MT, ML and MDS QSLs to the T-QSL. First we will consider  limiting cases of infinite temperature and trivial perturbation, and then turn to nontrivial perturbations and specific examples.

\medskip
\noindent  {\it Infinite temperature.}~~Consider a system at the infinite temperature, $\beta=0$ (here we assume that the Hilbert space is finite-dimensional). Trivially $\rho_t=\rho_0\sim \mathds{1}$ and $D_{\rm tr}(\rho_0,\rho_t)=0$ in this case. This result is readily reproduced by  T-QSL \eqref{T-QSL} as well as MDS QSL in its original form \eqref{MDS QSL original}. However, both MT \eqref{MT QSL} and ML \eqref{ML QSL} QSLs provide  loose $O(\sqrt{N})$ bounds.

\medskip
\noindent  {\it Trivial perturbation.}~~The perturbation is called {\it trivial} if $[H_0,V]=0$. For a trivial perturbation the dynamics is also trivial, $\rho_t=\rho_0$, and the performance of  QSLs is absolutely analogous to the infinite temperature case.
%, and  $D_{\rm tr}(\rho_0,\rho_t)=0$. The resulting performance of QSLs is absolutely analogous to the previous case: T-QSL  and MDS QSL in its original form are exact while MT and ML QSLs provide  loose $O(\sqrt{N})$ bounds.

\medskip
\noindent  {\it Local perturbation.}~~ We refer to $V$ as a {\it local} perturbation whenever $\langle V^2 \rangle_\beta$ is finite in the thermodynamic limit. If the perturbation is local, the MDS bound \eqref{MDS QSL} is finite in the thermodynamic limit.

\medskip
\noindent  {\it Finitely disturbing perturbation.}~~ We refer to $V$ as a {\it finitely disturbing} perturbation whenever  $\langle [H_0,V]^2 \rangle_\beta$ is finite in the thermodynamic limit. The T-QSL \eqref{T-QSL} is finite in the thermodynamic limit whenever the perturbation is finitely-disturbing.

For most physical Hamiltonians $H_0$ (in particular, for lattice Hamiltonians with short-range interactions) a local perturbation is also a finitely disturbing one. The opposite, however, is not necessarily true, as will be shown in an example below. Therefore a T-QSL is expected to outperform the MDS QSL whenever the perturbation is finitely disturbing but not local.

\medskip
\noindent  {\it Spin-boson model.}~~ As an explicit but still quite general example, we consider a spin-boson model
\be\label{H spin-boson}
H_0=\Omega\, \sigma^z+\frac1{\sqrt{N}}\,\sigma^x \sum_k g_k(a_k^\dagger+a_k)+ \sum_k \omega_k\,a_k^\dagger a_k.
\ee
Here $\sigma^{x,z}$ are Pauli matrices of a two-level system that is coupled to $N$ bosonic modes (oscillators) $a_k$. With the appropriate choices of the energies $\Omega, \omega_k$ and coupling constants $g_k$, this Hamiltonian describes a multitude of many-body systems \cite{Leggett_1987_spin-boson_review}. Note that  a prefactor $1/\sqrt N$ in the interaction term has been explicitly singled out, so that  coupling constants $g_k$ remain independent on the system size~\cite{Leggett_1987_spin-boson_review}.

We will consider two types of perturbations in the spin-boson model.  The first one is local and affects the spin degree of freedom with $V=\varepsilon\sigma^x$. Both $\Delta E$ and $\sqrt{\overline E}$ are dominated by the last term in the Hamiltonian \eqref{H spin-boson} and diverge as $\sqrt{N}$, the same divergence plaguing the MT and ML QSLs, in accordance with the general considerations (the same is true for another perturbation considered below). The MDS bound~\eqref{MDS QSL} reads $D_{\rm tr}(\rho_0,\rho_t)\leq \sqrt{2}\,  \varepsilon\, t$. The T-QSL produces a somewhat different bound, $D_{\rm tr}(\rho_0,\rho_t)\leq \sqrt{2\sqrt2 \, \varepsilon\,\Omega\,\beta\, t}$. Both the MDS QSL and T-QSL avoid a spurious divergence in the thermodynamic limit.

Another perturbation we consider shifts the energies of oscillators:
\be
V=\sum_k \delta\omega \, a_k^\dagger a_k.
\ee
This perturbations is finitely disturbing, but not local. The MDS bound \eqref{MDS QSL} now diverges as $N$, namely $D_{\rm tr}(\rho_0,\rho_t)\leq \sqrt2\, \delta\omega \,t\, \overline n_\beta\, N$, where $\overline n_\beta \equiv \sum_k \langle a^\dagger_k a_k \rangle_\beta/N$ is the average number of excitation per mode, with $\langle a^\dagger_k a_k \rangle_\beta=1/(e^{\beta \omega_k}-1)$ being the Bose-Einstein distribution (here and in what follows the subleading in $N$ terms  are omitted from the bounds).  However, this divergence is spurious: the T-QSL provides a finite bound
\be\label{T-QSL spin-boson}
D_{\rm tr}(\rho_0,\rho_t) \leq \sqrt{\delta\omega\,\widetilde{g}\,\beta\, t} \, \sqrt[4]{2(1+2\,\widetilde{n}_\beta)},
\ee
where $\widetilde{g}^2\equiv \sum_k g_k^2/N$ and $\widetilde{n}_\beta\equiv\sum_k g_k^2 \langle a^\dagger_k a_k \rangle_\beta /\sum_k g_k^2$ are finite in the thermodynamic limit.
Thus, the T-QSL is the only reasonable bound in this case.

\medskip
\noindent  {\it Mobile impurity model.}~~ As a second example, we consider a  model describing a single mobile impurity particle with mass $m$ immersed in a fluid. The Hamiltonian reads
\be
H_0=H_{\rm f}+P^2/(2m) + H_{\rm imp-f},
\ee
where $H_{\rm f}$ is the Hamiltonian of the fluid, $P$ is the momentum of the impurity and $H_{\rm imp-f}$ describes the interaction between the impurity and the fluid. For the ease of notations, we consider a one-dimensional case. The fluid and the impurity are in a box with the size $L$, the number of the particles of the fluid is $N$ and the particle density $n=N/L$ is kept constant in the thermodynamic limit $N,L\rightarrow\infty$. The perturbation reads
\be
V=F X,
\ee
where $X\in[0,L]$ is the coordinate of the impurity. This perturbation describes a force $F$ applied to the impurity at time $t=0$. This or similar setups are actively studied theoretically  \cite{Gangardt2009,schecter2012dynamics,Gamayun2014kinetic,Gamayun2014keldysh,lychkovskiy2015perpetual,schecter2016quantum,Castelnovo_2016_Driven,lychkovskiy2018necessary,Petkovic_2016_Dynamics,Petkovic_2020_Microscopic} and experimentally \cite{palzer2009quantum,catani2012quantum,meinert2016bloch}.

Again, the MT and ML bound produce a spurious $\sqrt{N}$ divergence. An even worse divergence plagues the MDS bound~\eqref{MDS QSL} which reads $ D_{\rm tr}(\rho_0,\rho_t)\leq \sqrt{2/3}\, N F  t/n $.
%, since $\sqrt{\langle (F X)^2 \rangle_\beta}=FL/\sqrt3$ up to $O(1)$ corrections.
In contrast, the T-QSL \eqref{T-QSL} is finite in the thermodynamic limit:
\be
D_{\rm tr}(\rho_0,\rho_t)\leq \sqrt{\beta t} \sqrt{(F/m)\sqrt{2 \langle P^2 \rangle_\beta}}.
\ee
Here the thermal average $\langle P^2 \rangle_\beta$ depends on the explicit form of $H_{\rm f}$ and $H_{\rm imp-f}$ and in each particular case can be calculated approximately or, for an integrable $H_0$, exactly (see e.g. \cite{Lychkovskiy_2018_quantum,gamayun2020zero}).\footnote{ Let us emphasise here that in the latter case $H_0+V$ is still nonintegrable and the calculation of $\rho_t$ is unfeasible.}

\medskip
\noindent  {\it T-QSL for a general driving.}~~Finally, let us state and prove a quantum speed limit for a Hamiltonian with an arbitrary time dependence. Namely, we consider a state $\rho_t$ initialized in the thermal state \eqref{thermal state} and evolving according to the von Neumann equation
\be\label{Schrodiger equation}
i\partial_t \rho_t = [H_0+V_t,\rho_t],
\ee
where $V_t$ is an arbitrary time-dependent perturbation. Then the thermal QSL reads
\be\label{T-QSL general}
D_{\rm tr}(\rho_0,\rho_t)\leq \sqrt{\beta \, \int_0^t dt' \sqrt{-2\,\langle [H_0,V_{t'}]^2 \rangle_\beta }}.
\ee
Essentially the same result cast in terms of the Bures angle reads
\be\label{T-QSL Bures angle general}
\mathcal L (\rho_0,\rho_t)\leq \arcsin \sqrt{\beta \, \int_0^t dt' \sqrt{-2\,\langle [H_0,V_{t'}]^2 \rangle_\beta }}.
\ee
In the particular case of time-independent $V_t=V$ the inequalities \eqref{T-QSL general} and \eqref{T-QSL Bures angle general} entail the inequalities  \eqref{T-QSL} and \eqref{T-QSL Bures angle}, respectively.

The proof of eqs. \eqref{T-QSL general},\eqref{T-QSL Bures angle general} goes as follows.
We will first bound
\be
D_t\equiv D(\rho_0,\rho_t),
\ee
where $D(\rho_0,\rho_t)$ is defined by eq. \eqref{D definition}.
Note that $\sqrt{\rho_t}$ also satisfies the  von Neumann equation, $i\partial_t \sqrt{\rho_t} = [H_0+V_t,\sqrt{\rho_t}].$ Therefore
\be
\partial_t D_t=i \, \tr([\sqrt{\rho_0}, V_t] \sqrt{\rho_t}).
\ee
We rewrite this equality in the eigenbasis of $H_0$:
\be\label{dtD}
 \partial_t D_t=\frac{i\beta}{2\sqrt{Z_0}}  \,\sum_{n,\, k} f_{E_nE_k}^\beta  \, \langle n| V_t |k \rangle \,(E_n-E_k)\,  \langle k | \sqrt{\rho_t} |n\rangle,
\ee
Here $|n\rangle$, $|k\rangle$ are eigenstates of $H_0$, $E_n$ and $E_k$ are corresponding eigenenergies, and we have defined the function of three variables
\be
f_{E\,E'}^\beta = \frac{e^{-\frac{\beta}{2}E}-e^{-\frac{\beta}{2}E'}}{\beta(E-E')/2}.
\ee
Then we  integrate eq.\eqref{dtD} over time,
\be\label{D=int}
D_t=\frac{i\beta}{2\sqrt{Z_0}} \int_0^t dt'  \sum_{n,\, k}f_{E_nE_k}^\beta\langle n| [H_0,V_{t'}] |k \rangle    \langle k | \sqrt{\rho_{t'}} |n\rangle,
\ee
where the equality  $ \langle n| V_{t'} |k \rangle(E_n-E_k)=\langle n| [H_0,V_{t'}] |k \rangle$ has been used.

Next, we apply the Cauchy-Bunyakovsky-Schwarz inequality to the r.h.s. of eq. \eqref{D=int}:
\begin{align}
D_t \leq & \frac{\beta}2
\int_0^t dt' \left( \sum_{n,\, k}Z_0^{-1} \left(f_{E_nE_k}^\beta\right)^2 \, \Big| \langle n| [H_0,V_{t'}] |k \rangle\Big|^2 \right)^{1/2}
\nonumber\\
&
\times \left( \sum_{n,\, k} \Big|\langle k | \sqrt{\rho_{t'}} |n\rangle\Big|^2 \right)^{1/2}.
\end{align}
The term in the second bracket reduces to $\tr \rho_{t'}=1$. The term in the first bracket can be estimated by using the inequality~\cite{il'in2020adiabatic}
\be
 \left(f_{E\,E'}^\beta\right)^2\leq e^{-\beta E}+e^{-\beta E'}
\ee
valid for any real $E$, $E'$ and $\beta$. This inequality after some basic algebra leads to the conclusion that  the first bracket is bounded from above by $-2\langle [H_0,V_{t'}]^2 \rangle_\beta.$ This way we obtain
\begin{align}
D_t \leq & \frac{\beta}2
\int_0^t dt' \sqrt{ -2\langle [H_0,V_{t'}]^2 \rangle_\beta }.
\end{align}
Combining this bound with (the looser versions of) inequalities \eqref{Holevo inequality} and \eqref{Audenaert inequality}  concludes the proof. $\square$

As is clear from the proof, the bounds \eqref{T-QSL general},\eqref{T-QSL Bures angle general} can be improved: one can,  first, use the tighter versions of inequalities \eqref{Holevo inequality}, \eqref{Audenaert inequality}, and, second, exploit the fact that $f_{E_nE_n}^\beta$ (observe identical subscripts) can be substituted by zero.  These straightforward but somewhat bulky improvements make the bound more tight quantitatively but, as far as we can see, do not bring new qualitative insights.

One can also attempt to optimise the obtained QSLs by applying a time-dependent gauge transformation to the Hamiltonian \cite{Wu_1993_new,pfeifer1995generalized,Gamayun_2021_Map,wu2020hamiltonian} prior to using the inequalities \eqref{T-QSL general},\eqref{T-QSL Bures angle general}. We leave this interesting research direction for future work.

We note that the definitions of trivial, local and finitely disturbing perturbations extend to the time-dependent $V_t$ without alterations, as well as the conclusions regarding the corresponding performance of QSLs.

\medskip
\noindent  {\it Summary.}~~ To summarize, we have proven  a quantum speed limit \eqref{T-QSL general} for a system prepared in a thermal state   of an initial Hamiltonian $H_0$ and evolving under a different, possibly time-dependent Hamiltonian $H_0+V_t$. By narrowing the set of initial states to thermal states only, we have enhanced the performance of this quantum speed limit (referred to as T-QSL) as compared to QSLs valid universally. We have compared  the T-QSL to three other QSLs, including the paradigmatic Mandelstam-Tamm and Margolus-Levitin ones, with the results summarised in Table \ref{table}. The superiority of the T-QSL is most spectacularly manifested for a class of nonlocal but locally disturbing perturbations $V_t$: To the best of our knowledge, in this class the T-QSL is the only QSL  providing a reasonable bound that avoids a spurious divergence in the thermodynamic limit. We have demonstrated this advantage explicitly for two exemplary systems -- a spin-boson and a mobile impurity models.

As a final remark, we stress that the T-QSL is capable to meaningfully bound the distance between {\it many-body} mixed states $\rho_t$ and $\rho_0$. This should be distinguished from the open system setup where one is interested in a {\it reduced} state of a few-level system (e.g. a qubit) coupled to a thermal bath, see e.g. \cite{deffner2013quantum,taddei2013quantum,Lapert_2013_understanding,Mukherjee_2015}. Since the trace distance is contractive with respect to taking partial trace, the T-QSL implies a bound on the distance between the reduced states, but not vice versa. For example, in the spin-boson model the right hand side of the inequality  \eqref{T-QSL spin-boson} bounds also  $D_{\rm tr}(\rho^s_0,\rho^s_t)$, where $\rho^s_t$ is the reduced density matrix of the spin.

\smallskip
\noindent  {\it Note added.} Very recently, a quantum speed limit for projections on linear subspaces has been reported  \cite{albeverio2020quantum}. This result can be applied to microcanonical thermal states, which nicely complements our results on the canonical thermal states.

\smallskip

\begin{acknowledgements}
\noindent  {\it Acknowledgements.} We are grateful to  A.V. Aristova for numerous fruitful discussions and to A.S. Holevo and A.E. Teretenkov for valuable remarks. The work was supported by the Russian Science Foundation under the grant N$^{\rm o}$ 17-71-20158.
\end{acknowledgements}

\bibliography{C:/D/Work/QM/JabRef/bibliography,C:/D/Work/QM/Bibs/LZ_and_adiabaticity,C:/D/Work/QM/Bibs/AQC,C:/D/Work/QM/Bibs/QIP,C:/D/Work/QM/Bibs/QSL,C:/D/Work/QM/Bibs/1D,C:/D/Work/QM/Bibs/dynamically_integrable}

\end{document}